\documentclass[reprint,
bibnotes,
amsmath,amssymb,
aps,
]{revtex4-2}
\usepackage[justification=centerlast]{caption}
\usepackage{subcaption}
\usepackage{graphicx}% Include figure files
\usepackage{dcolumn}% Align table columns on decimal point
\usepackage{float}%to remove spaces before or after figure
\usepackage{bm}% bold math
\begin{document}
%\preprint{APS/123-QED}
\title{Engineering of Chern number of topological bands in bilayer graphene by in-plane magnetic field and electrical bias}% Force line breaks with \\

\author{Narjes Kheirabadi}%\email{n.kheirabadi@Lancaster.ac.uk}
\affiliation{Department of Physics, Sharif University of Technology, Tehran 14588-89694, Iran}%
%\affiliation{$^2$ School of Physics, Institute for Research in Fundamental Sciences (IPM), P.O.Box:19395-5531, Tehran, Iran}%

%\date{\today}% It is always \today, today,
             %  but any date may be explicitly specified

\begin{abstract}
Based on the full Hamiltonian of bilayer graphene, phase transitions are realized by the change of the in--plane magnetic field and the electrical bias in bilayer graphene. We show that the engineering of Chern numbers of four bands is possible by an applied in--plane magnetic field and an electrical bias in bilayer graphene. Our results are promising for the exploration of new topological phenomena in 2D materials.
\end{abstract}

%\pacs{Valid PACS appear here}% PACS, the Physics and Astronomy
                             % Classification Scheme.
%\keywords{Suggested keywords}%Use showkeys class option if keyword
                              %display desired
\maketitle
\section{Introduction}

A Chern insulator is a 2--dimensional material that holds chiral edge states caused by the combination of topology with time--reversal symmetry breaking \cite{ovchinnikov2022topological} with a non--zero Chern number whose topological phases can be characterized by this number.

In this paper, we aim to determine the influence of the in-plane magnetic field on the topology of 4-bands based on the full Hamiltonian of bilayer graphene in a steady parallel magnetic field. To do so, we have calculated the Chern number for 4-bands where a steady in--plane magnetic field has broken time--reversal symmetry in a 2--dimensional material, bilayer graphene, in the presence and absence of an electrical bias. We show that in the absence of an electrical bias, bilayer graphene under a steady in--plane magnetic field is a platform to discover Chern insulator physics caused by the orbital effect of electrons under an in--plane magnetic field. 

The influence of an in-plane field on bilayer graphene has been considered previously \cite{mariani2012fictitious, pershoguba2010energy, roy2013bilayer, van2016transport, mucha2011strained, son2011electronic}, but simplified tight--binding models were used neglecting small
tight--binding parameters such as 
skew interlayer coupling, and interlayer potential asymmetry. In this article, we show that these parameters are essential to have a correct understanding of the topological properties of bilayer graphene under a parallel magnetic field. 
 
Additionally, a comparison between the result of this article which is based on a 4--band model tight-- binding Hamiltonian and the effective theory and the 2--band model based on the effective Hamiltonian studied in Ref.~\cite{kheirabadi2016magnetic} shows that the effective theory and the low--energy Hamiltonian are not appropriate to study the Chern number of electronic bands. For example, when we consider the interlayer asymmetry, its effect could not be understood well based on the effective Hamiltonian and even the low--energy Hamiltonian. According to the effective Hamiltonian, the effect of interlayer asymmetry is opening a gap between low energy bands and this gap could be closed by the in-plane magnetic field. In this work, we try to have a better understanding of the effect of this factor on the topological properties of bilayer graphene under a parallel magnetic field by the calculation of the Chern number of  four bands and an analysis based on the change of the band gaps related to the applied magnetic field. 

\section{Hamiltonian and Calculation method}
\begin{figure}
   \centering   
   \includegraphics[scale=0.57]{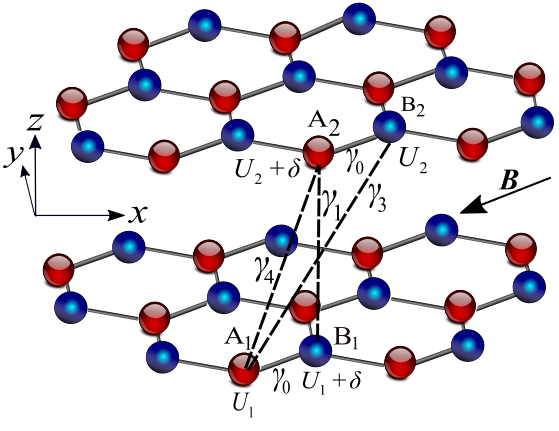}
   \caption{The AB--stacked bilayer graphene unit cell. $A_1$ and $B_1$ atoms on the bottom layer and $A_2$ and $B_2$  on the top layer have been depicted. Straight lines point out intralayer coupling $\gamma_0$, vertical dashed lines also show interlayer coupling $\gamma_1$ and skew interlayer couplings $\gamma_3$ and $\gamma_4$. Parameters $U_1$ and $U_2$ indicate different on--site energies where $\delta$ is the dimer sites asymmetry related parameter.} 
    \label{ab}
\end{figure}

The AB--stacked bilayer graphene structure and its related parameters with a lattice constant $a$ and interlayer distance $d$ is depicted in the Fig.~\ref{ab}. According to this figure, different on--site energies are $U_1$ and $U_2$ on the $A_1$ and $B_2$ sites, respectively and those are the on--site energies of the two layers. $\delta$ is also an energy difference between $A$ and $B$ sites on each layer.  
Based on the tight--binding Hamiltonian and the  nearest neighbour approximation with hopping parameters $\gamma_0, \gamma_1, \gamma_3, \gamma_4$  and on--site energies stated in Fig.~\ref{ab},
to calculate the Chern number of each band, we have used the method introduced in Ref.~\cite{fukui2005chern} and the Code in Ref.~\cite{ning}. To do so, we have written Hamiltonian of the nearest-neighbour model, previously described in \cite{kheirabadi2018electronic, kheirabadi2016magnetic}, based on $k_1$ and $k_2$ where each wave vector $\mathbf{k}$ is considered as $\mathbf{k}=\mathbf{b}_i \frac{k_i}{2 \pi}$, $\mathbf{b}_i$ are reciprocal lattice unit vectors \cite{fukui2005chern}. Hence, the Hamiltonian of bilayer graphene under a parallel magnetic field based on $k_1$ and $k_2$ has the following form ($a=1$)
\begin{eqnarray}\label{Hamiltonian}
H=\begin{pmatrix}
U_1 &-\gamma_0 f_1 &\gamma_4 f &-\gamma_3 f_3 \\ 
-\gamma_0 f_1^* &U_1+\delta &\gamma_1 &\gamma_4 f\\ 
\gamma_4 f^* &\gamma_1  &U_2+\delta &-\gamma_0 f_2\\ 
-\gamma_3 f_3^* &\gamma_4 f^* &-\gamma_0 f_2^* & U_2
\end{pmatrix}
\end{eqnarray}
where
\begin{eqnarray}
f&&=1+ \exp(-i k_1)+ \exp(-i(k_1+k_2)),\\
f_1&&=\exp\big(-i (k_1+k_2) - i e \frac{ d }{2 \sqrt{3} \hbar} B_x\big)\nonumber\\
&& + \exp\big(- i k_1 + i e \frac{ d }{2 \hbar} \big( \frac{B_y}{2}+\frac{B_x}{2 \sqrt{3}}\big)\big)\nonumber\\
&& + \exp\big(- i e \frac{ d }{2 \hbar} \big( \frac{ B_y}{2}-\frac{B_x}{2 \sqrt{3}}\big)\big),\\
f_2&&=\exp\big(-i (k_1+k_2) + i e \frac{ d }{2 \sqrt{3} \hbar} B_x\big)\nonumber\\
&& + \exp\big(- i k_1 - i e \frac{ d }{2 \hbar} \big( \frac{B_y}{2}+\frac{B_x}{2 \sqrt{3}}\big)\big)\nonumber\\
&& + \exp\big( i e \frac{ d }{2 \hbar} \big( \frac{ B_y}{2}-\frac{B_x}{2 \sqrt{3}}\big)\big),\\
f_3&&=\exp(-i k_1)+ \exp(-i (k_1+k_2))+ \exp(-i(2 k_1+k_2)).\nonumber\\
\end{eqnarray} 
Here, $e$, $e>0$, is the electron charge, $\mathbf{B}=(B_x, B_y)$ is the applied in--plane magnetic field. In all of the calculations, numerical amounts are based on Ref.~\cite{kheirabadi2016magnetic} and those are $\gamma_0=3.16$ eV, $\gamma_1=0.39$ eV, $\gamma_3=0.38$ eV, $\gamma_4=0.14$ eV, $\delta=0.2$ eV and $\Delta= U_2-U_1$. In bilayer graphene, the presence of skew interlayer coupling $\gamma_3$ causes a distortion of the Fermi circle, known as trigonal warping \cite{dresselhaus1981intercalation, ando1998berry, mccann2006landau}. At very low energy,
this warping leads to a breaking of the Fermi surface into four different pockets. Also, $\delta$ together with $\gamma_4$ breaks the particle-hole symmetry \cite{moulsdale2020engineering}.

When the Fermi energy states in a gap, the Hall conductance is $\sigma_{xy}=-e^2/h \Sigma_{n} c_n$, where $c_n$ is the Chern number of the $n$th Bloch band, and the sum is over the filled bands \cite{fukui2005chern, xiao2010berry}. The Chern number of $n$th band is defined as \cite{fukui2005chern, xiao2010berry}
\begin{eqnarray}
c_n=\frac{1}{2\pi}\int_{BZ} \mathbf{F}_n(\mathbf{k}) d\mathbf{k},
\end{eqnarray}
where $\mathbf{F}_n(\mathbf{k})$ is the Berry curvature.

To continue, we study the effect of the electrical bias on the topology of bands by the calculation of the Chern number of four bands in bilayer graphene under a steady in--plane magnetic field based on the method introduced in Ref.~\cite{fukui2005chern}.

\section{Discussion}
\subsection{$\Delta\neq 0$}
\begin{figure}[H]
   \centering   
   \includegraphics[width=0.5\textwidth]{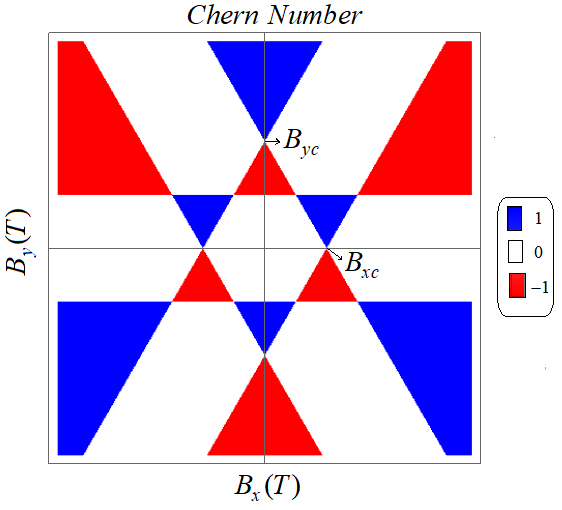}
   \caption{The phase diagram of AB--stacked bilayer graphene under the steady in--plane magnetic field for the lowest conduction band. The accuracy of the calculation is $0.25$ T.} 
    \label{PD}
\end{figure}

\begin{figure}[H]
   \centering   
   \includegraphics[width=0.5\textwidth]{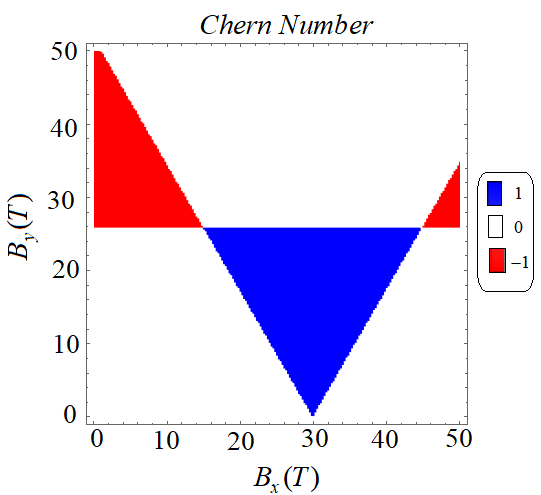}
   \caption{The phase diagram of AB--stacked bilayer graphene under a steady in--plane magnetic field for the lowest conduction band where $\Delta=10$ m eV. The accuracy of the calculation is $0.25$ T.} 
    \label{PD2}
\end{figure}

\begin{figure}[H]
   \centering   
   \includegraphics[width=0.50\textwidth]{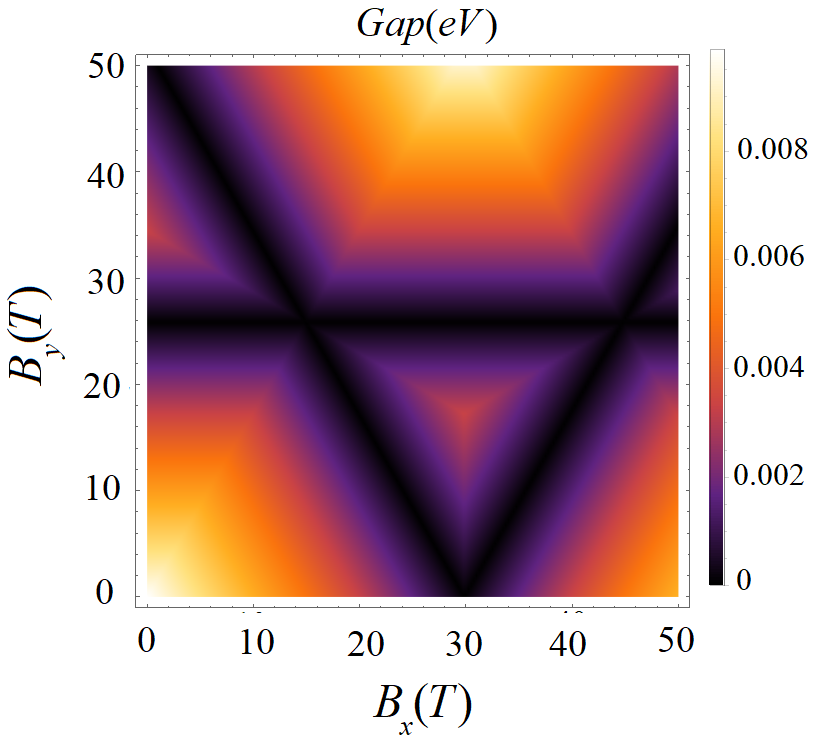}
   \caption{The diagram of the change of the gap between the lowest conduction band and the band above it by the change of the magnetic field where $\Delta=10$ m eV. The accuracy of the calculation is $0.25$ T.} 
    \label{Gap2}
\end{figure}

Based on Ref.~\cite{fukui2005chern}, for a $400 \times 400$ lattice Brillouin zone, we have derived the phase diagram of the gaped bilayer graphene under an in--plane magnetic field when spatial inversion symmetry is broken in understudy bilayer graphene by an applied gate voltage (Fig.~\ref{PD}).
However, near phase transition points, where the gap closes or is near zero, a $700 \times 700$ lattice Brillouin zone has also been used.  According to this numerical calculations, $B_{yc}$ and $B_{xc}$ parameters in Fig.~\ref{PD} obey the following equations:
\begin{eqnarray}
B_{yc}&&\simeq  \frac{\Delta(meV)}{10} \times 51.704,\nonumber\\
B_{xc}&&\simeq \frac{B_{yc}}{\sqrt{3}}.
\end{eqnarray}
Furthermore, the Chern numbers of the valence bands are equal to zero. These calculations show that when the symmetry is broken by an electrical bias, the gap between the highest valence band and the first conduction band does not close by the change of the in--plane magnetic field, and band crossing does not happen for these two bands. For instance, if the on--site energies change so that  $U_2=-U1= \Delta/2$ and $\Delta=10$ m eV, we will have the phase diagram depicted in Fig.~\ref{PD2} for $0 \le B_x \le 50$ T and $0 \le B_y \le 50$ T.  
In addition, we have calculated the electronic gap between the lowest conduction band and the band above it (Fig.~\ref{Gap2}) based on the Hamiltonian of bilayer graphene in the parallel magnetic field in Ref.~\cite{kheirabadi2022quantum}. 
A comparison between Fig.~\ref{PD2} and Fig.~\ref{Gap2} shows that the gap closing areas and the phase transition areas are the same area. 

\begin{figure*}
   \centering   
   \includegraphics[width=1\textwidth]{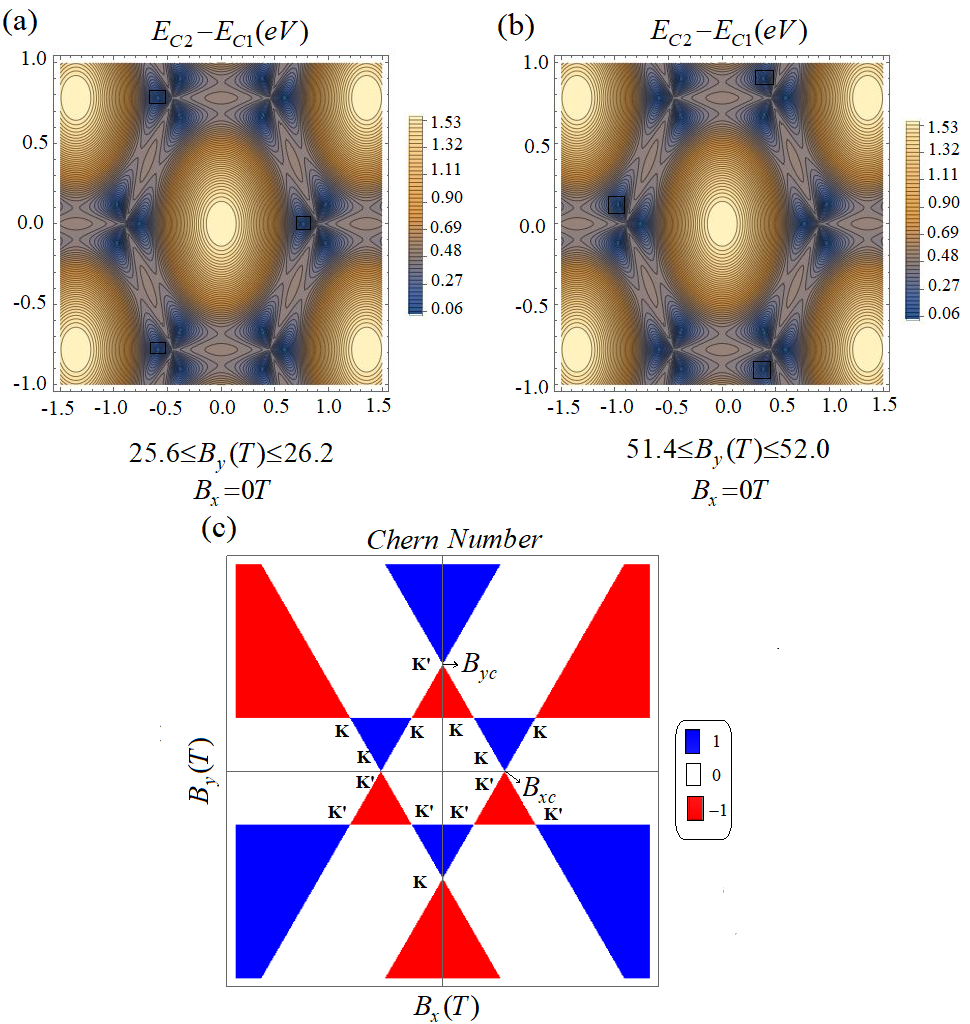}
   \caption{(a and b) Contour plots of the change of the gap between the lowest conduction band and the band above it for bilayer graphene under a parallel magnetic field where $B_x=0$ T and $B_y$ changes near phase transition points for (a) $B_y$ near $B_{yc}/2$ and (b) for $B_y$ near $B_{yc}$; the minimum of the gap is a point in one of the square areas depicted on the diagram ($\Delta = 10$ m eV) (c) The phase diagram of bilayer graphene under a parallel magnetic field regarding the gap closing position. In this figure, it has been depicted that the gap closing is near $K$ or $K'$ valley in the phase transition points. } 
    \label{Valley2}
\end{figure*}

Moreover, when time reversal symmetry is valid in bilayer graphene, opposite valleys have opposite Chern numbers. When bilayer graphene is under an in--plane magnetic field, time--reversal symmetry is broken and consequently the symmetry between two valleys is also broken. 
For instance, according to Fig.~\ref{Valley2} (a), when $B_x=0$ T, near the phase transition point $B_{yc}/2$ (Fig.~\ref{PD}), where the magnetic field in the $y$ direction changes by $0.1$ T steps from $25.6$ T to $26.2$ T, the gap closing points are located near the $K$ valley. However, when the magnetic field in $y$ direction changes by $0.1$ T steps from $51.4$ T to $52.0$ T, the electronic gap between the conduction band and the band above it increases near $K$ points and decreases near $K^\prime$ points (Fig.~\ref{Valley2} (b)); these magnetic fields are near to $B_{yc}$ transition point (Fig.~\ref{PD}). Then, by increasing $B_y$, the gap opens near $K$ and $K'$ points. Consequently, a valley selectivity effect is predictable where the gap closes between the lowest conduction band and the band above it (Fig.~\ref{Valley2} (c)).

\begin{figure*}
   \centering   
   \includegraphics[width=1.0\textwidth]{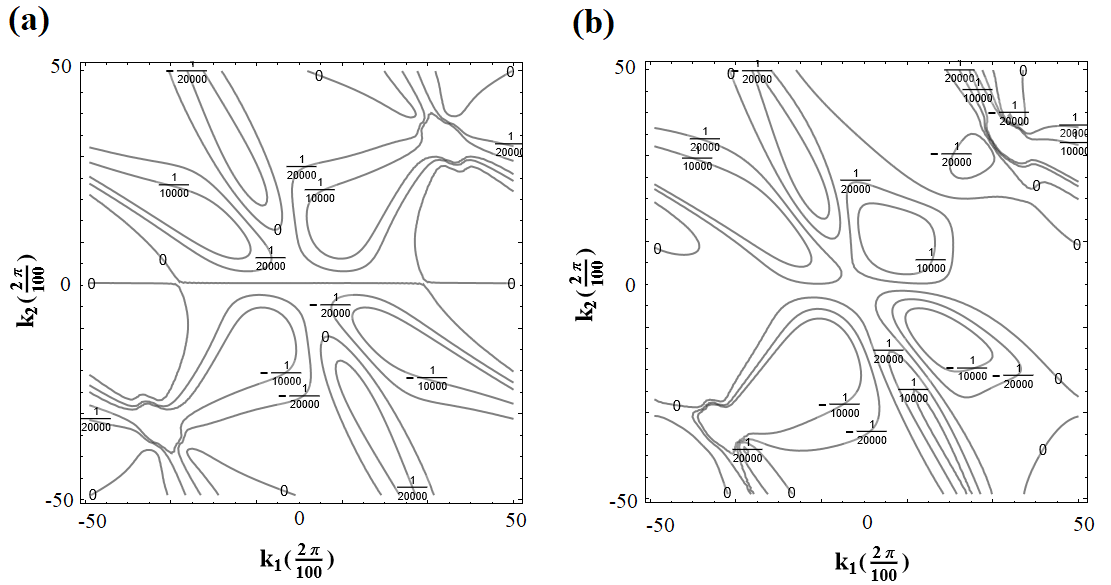}
   \caption{Isovalue contours for Berry curvature distributions of the highest valence band for a $100 \times 100$ lattice Brillouin zone of bilayer graphene under $\Delta = 10$ m eV bias where (a) applied magnetic field is zero (b) $B_x=30$ T and $B_y=-30$ T. Figure determined with
numerical calculations as described in the main text. Constant Berry curvature contours in the $0$, $\pm 1/20000$, and $\pm 1/10000$ (a.u) are plotted. }\label{0}
\end{figure*}

However, the Hall conductance is zero near the Fermi energy and this system is topologically trivial, even if the valence band Chern number is zero, the valence band Berry curvature is not zero everywhere in the momentum space and the in-plane magnetic field has broken the inversion symmetry in the Berry curvature over the Brillouin zone (Fig.~\ref{0}). Additionally, the Chern number spectral function is like a density of states of Berry curvature, so the regions of positive Berry curvature in momentum space will contribute to a positive Chern number spectral function at some frequencies, and the regions of negative Berry curvature in momentum space will contribute to a negative Chern number spectral function in some other frequencies; the Chern number spectral function times frequency is what is measured in the circular dichroism experiment.  Hence, the whole system is topologically trivial because of the zero Chern number of the valence band, but the valence band Berry curvature is not zero everywhere in the momentum space. As a result, the system will still have circular dichroism as described in Ref.~\cite{molignini2023probing}.

According to the method introduced in Ref.~\citep{fukui2005chern}, we can show that dependent on the sign of $\Delta$, the Berry curvature has a maximum near to $K$ and a minimum near to $K^\prime$ or inverse. Because the valley Chern number is the sum of Berry curvature over a large region of k space around one valley, different valleys have opposite Chern numbers. And, a valley Chern insulator is predictable here and the sign of each valley Chern number could be reversed by the change of the sign of $\Delta$ factor. This effect is similar to previous results for gaped bilayer graphene \cite{PhysRevLett.106.156801, RevModPhys.82.1959}. However, the in--plane magnetic field has broken the inversion symmetry of the Berry curvature over the Brillouin zone (Fig.~\ref{0}), it does not change the Chern number of each valley.

We can show that considering $\delta$ and $\gamma_4$ is essential to have correct Chern numbers for 4-bands. In fact, ignoring the interlayer asymmetry potential, $\delta$, results in different values for $B_{xc}$ and $B_{yc}$  (Fig.~\ref{PD}), and ignoring the $\gamma_4$ hopping factor results in different Chern numbers for each band. If we do not consider $\gamma_4$ hopping parameter, topological properties of bilayer graphene under an in--plane magnetic field will be predicted incorrectly, for example for some amount of magnetic field, a non-zero Chern number for the valence band will be predicted.   
\subsection{$\Delta = 0$}
\begin{figure*}
   \centering   
   \includegraphics[width=0.99\textwidth]{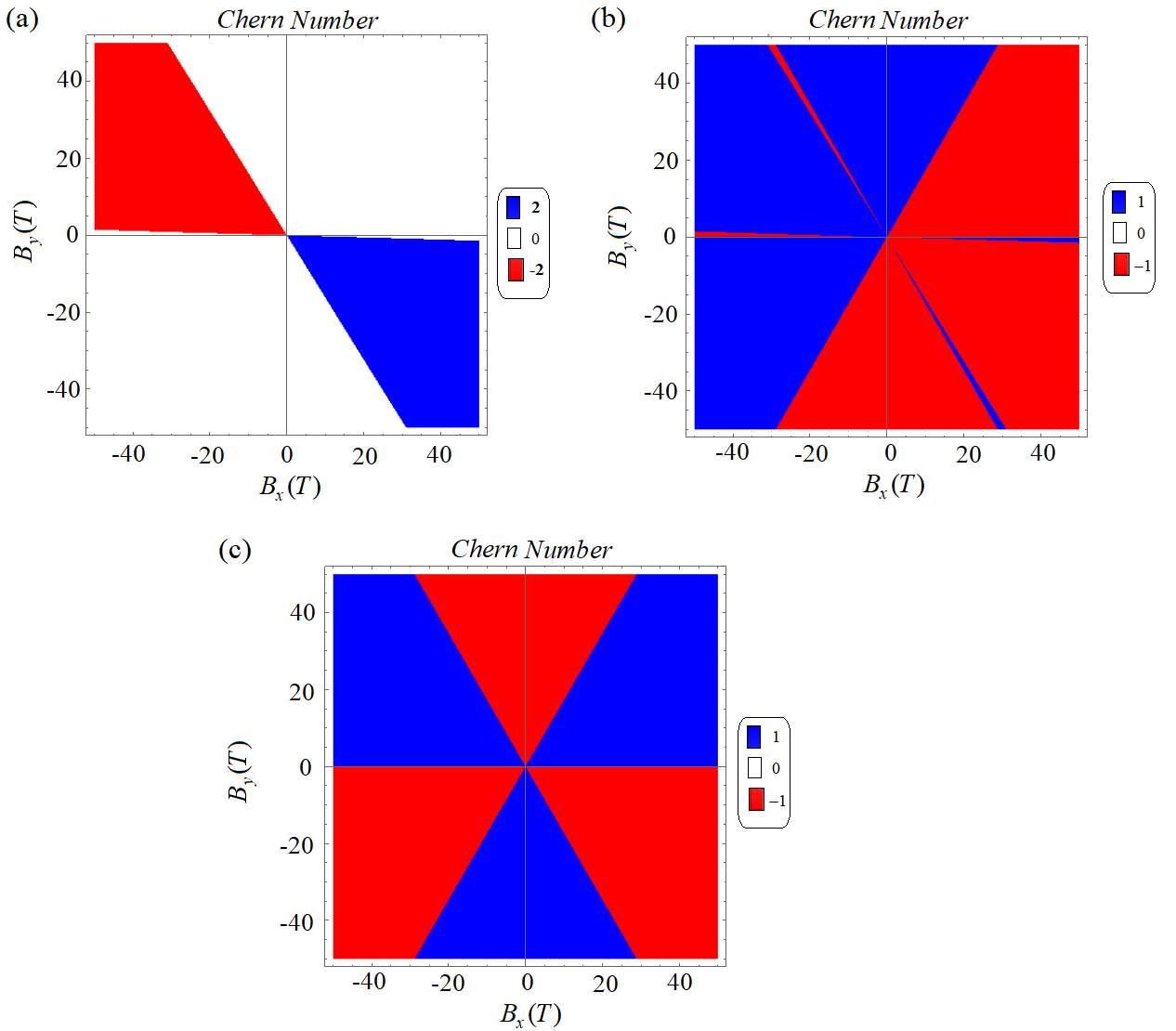}
   \caption{The phase diagram of bilayer graphene under a parallel magnetic field in the absence of the electrical bias for (a) the highest valence band, (b) the lowest conduction band and the band above it (c).} 
    \label{Gap0}
\end{figure*}

In the second case, the phase diagram of bilayer graphene under an in--plane magnetic field in the absence of any electrical bias is depicted in Fig.~\ref{Gap0}.
According to Fig.~\ref{Gap0}, when there is not any applied electrical bias on bilayer graphene under an in--plane magnetic field, bilayer graphene is a Chern insulator, and the Chern number is dependent on the magnetic field direction and strength. Because a band with a non-zero Chern number is topologically non-trivial, the highest occupied band is non-trivial and completely filled, so the state is called a topological insulator, and chiral edge states and quantized longitudinal conductivity are predictable in this system. 

According to Fig.~\ref{Gap0}, the phase diagram of bilayer graphene under a parallel magnetic field is different from what has been predicted based on low energy Hamiltonian and effective Hamiltonian described in Ref.~\citep{kheirabadi2018electronic}; so, the 4--band model is necessary to have a correct understanding about the topological properties of bilayer graphene under an in--plane magnetic field.
\section{Conclusion}
In the presence of an electrical bias, the gaped bilayer graphene in the presence of an applied magnetic field is a valley Chern insulator. The broken time--reversal symmetry by the applied magnetic field has broken the symmetry between different valleys and it leads to a valley selectivity in the gaped bilayer graphene where the gap closes between the lowest conduction band and the band above it. In the absence of an electrical bias, bilayer graphene under a steady in--plane magnetic field is a Chern insulator. In this Chern insulator the orbital effect of the in-plane magnetic field tunes topological bands properties and can be manipulated by the magnetic field direction and strength. In this case, in the edge states of bilayer graphene under an in-plane magnetic field, the electron movement is ballistic, and those are topologically protected and robust, and an electron disperses linearly, as high-energy physics massless particles.
\section{Acknowledgement}
The author thanks A.~Langari, T.~Fukui, E.~McCann and A.~Vaezi for useful discussions.
\bibliography{bib} 
\bibliographystyle{unsrtnat}
\end{document}